# Evidence of an odd-parity hidden order in a spin-orbit coupled correlated iridate

L. Zhao[1,2], D. H. Torchinsky[1,2], H. Chu[2,3], V. Ivanov[1], R. Lifshitz[1,4], R. Flint[5], T. Qi[6], G. Cao[6] & D. Hsieh[1,2]

[1]*Department of Physics, California Institute of Technology, Pasadena, CA 91125, USA.*

[2]*Institute for Quantum Information and Matter, California Institute of Technology, Pasadena, CA 91125, USA.*

[3]*Department of Applied Physics, California Institute of Technology, Pasadena, CA 91125, USA.*

[4]*Raymond and Beverly Sackler School of Physics and Astronomy, Tel Aviv University, Tel Aviv 69978, Israel.*

[5]*Department of Physics and Astronomy, Iowa State University, Ames, Iowa 50011, USA.*

[6]*Center for Advanced Materials, Department of Physics and Astronomy, University of Kentucky, Lexington, Kentucky 40506, USA.*

**A rare combination of strong spin-orbit coupling and electron-electron correlations makes the iridate Mott insulator $Sr_2IrO_4$ a promising host for novel electronic phases of matter[1,2]. The resemblance of its crystallographic, magnetic and electronic structures[1–6] to $La_2CuO_4$, as well as the emergence upon doping of a pseudogap region[7–9] and a low temperature $d$-wave gap[10,11], has particularly strengthened analogies to cuprate high-$T_c$ superconductors[12]. However, unlike the cuprate phase diagram that features a plethora of broken symmetry phases[13] in a pseudogap region that include charge density wave, stripe, nematic and possibly intra-unit cell loop-current orders, no broken symmetry phases proximate to the parent antiferromagnetic Mott insulating phase in $Sr_2IrO_4$ have been observed to date, making the comparison of iridate to cuprate phenomenology incomplete. Using optical second harmonic generation, we report evidence of a hidden non-dipolar magnetic order in $Sr_2IrO_4$ that breaks both the spatial inversion and rotational symmetries of the underlying tetragonal lattice. Four distinct domain types corresponding to discrete 90° rotated orientations of a pseudovector order parameter are identified using nonlinear optical microscopy, which is expected from an electronic phase that possesses the symmetries of a magneto-electric loop-current order[14–18]. The onset temperature of this phase is monotonically suppressed with bulk hole doping, albeit much more weakly than the Néel temperature, revealing an extended region of the phase diagram with purely hidden order. Driving this hidden phase to its quantum critical point may be a path to realizing superconductivity in $Sr_2IrO_4$.**

The crystal structure of the 5$d$ transition metal oxide $Sr_2IrO_4$ is tetragonal and centrosymmetric with four-fold ($C_4$) rotational symmetry about the $c$-axis[2–6]. It is composed of stacked $IrO_2$ square lattices whose unit cell is doubled relative to the $CuO_2$ square lattices found in high-$T_c$ cuprates owing to a staggered rotation of the octahedral oxygen cages (Fig. 1a). Despite the much larger spatial extent of 5$d$ versus 3$d$ orbitals, $Sr_2IrO_4$ is a Mott insulator by virtue of strong atomic spin-orbit coupling[1,2]. The electronic states near the Fermi level derive predominantly from the $t_{2g}$ $d$-orbitals of iridium ions, which are well approximated as two completely filled spin-orbital entangled $J_{eff} = 3/2$ bands and one half-filled $J_{eff} = 1/2$ band that splits into an upper and lower Hubbard band by an on-site Coulomb interaction. Below a Néel temperature $T_N \sim 230$ K, the spin-orbital entangled $J_{eff} = 1/2$ magnetic dipole moments undergo three-dimensional long-range ordering into an orthorhombic antiferromagnetic structure[2–5] that preserves global inversion symmetry but lowers the rotational symmetry of the system from $C_4$ to $C_2$. No additional symmetry breaking has otherwise been observed by neutron or x-ray diffraction.

Optical second harmonic generation (SHG), a frequency doubling of light through its nonlinear interaction with a material, is strongly affected by point group symmetry changes[19] and can be used to search for hidden phases with higher rank tensor order parameters that are difficult to detect using conventional probes[18,20]. It is particularly sensitive to inversion symmetry breaking because the leading order electric-dipole (ED) contribution to SHG, which is described by a third-rank susceptibility tensor $\chi_{ijk}^{ED}$ that relates the nonlinear polarization at the second harmonic frequency $2\omega$ to the incident electric field via $P_i(2\omega) \propto \chi_{ijk}^{ED} E_j(\omega) E_k(\omega)$, is only allowed if the crystal lacks an inversion center. Otherwise SHG can only arise from much weaker higher rank multipole processes as is the case for the $Sr_2IrO_4$ crystal structure described below. The rotational symmetry of a crystal can be directly determined by performing a rotational anisotropy (RA) SHG experiment where the intensity of obliquely reflected SHG is measured as a function of the angle $\varphi$ through which the scattering plane is rotated about the surface normal of the crystal (Fig. 1b). For our experiment, the incident light is focused onto optically flat regions of the crystal surface with a spot size $d < 100$ μm. The polarizations of the incident (in) and reflected (out) beams can be independently selected to be either parallel (P) or perpendicular (S) to the scattering plane, which allows for different nonlinear susceptibility tensor elements to be probed.

Figure 1c shows room temperature RA-SHG patterns of $Sr_2IrO_4$ measured by rotating the scattering plane about its *c*-axis. The data measured under both $P_{in}$-$P_{out}$ and $P_{in}$-$S_{out}$ polarization geometries (see Supplementary Information S1 for other geometries) exhibit a clear $C_4$ symmetry and are completely accounted for by the leading order non-local bulk contribution to SHG of electric-quadrupole (EQ) type[6], which can be written as an effective nonlinear polarization $P_i(2\omega) \propto \chi^{EQ}_{ijkl} E_j(\omega) \nabla_k E_l(\omega)$. No surface ED contribution to SHG was detected within our instrument sensitivity[6]. As discussed in Ref. [6], the rotation of the SHG intensity maxima away from the crystallographic *a* and *b* axes and their modulated amplitude in the $P_{in}$-$S_{out}$ pattern clearly signify the absence of mirror symmetry across the *ac* and *bc* planes. Expressions for the RA-SHG patterns (S2) calculated using the set of independent non-vanishing tensor elements of the fourth-rank susceptibility tensor $\chi^{EQ}_{ijkl}$ derived from the experimentally established centrosymmetric 4/*m* crystallographic point group of $Sr_2IrO_4$ fit the data extremely well (Fig. 1c).

Remarkably, the rotational symmetry of the RA-SHG patterns measured at low temperature ($T = 170$ K) is reduced from $C_4$ to $C_1$ (Fig. 1d). This is not caused by a structural distortion because extensive neutron and resonant x-ray diffraction studies show no change in crystallographic symmetry below room temperature[2–5]. It cannot be accounted for by the antiferromagnetic structure that develops below $T_N \sim 230$ K because that has a centrosymmetric orthorhombic magnetic point group (2/*m*1′) with $C_2$ symmetry (S3). Ferro- or antiferroelectric order can also be ruled out based on the absence of any anomaly in the temperature dependence of the dielectric constant of $Sr_2IrO_4$ above $T = 170$ K[21]. Instead, the ordering of a higher multipolar degree of freedom that coexists with the $J_{eff} = 1/2$ moment in each $IrO_6$ octahedron is left as the most plausible explanation. An ordering of higher rank parity-even magnetic multipoles with the same propagation wave vector as the antiferromagnetic structure has, in fact, been proposed to occur in $Sr_2IrO_4$ below $T_N$[22,23]. Such an order would naturally be difficult to detect using diffraction based probes because it preserves the translational symmetry of the antiferromagnetic lattice and imparts no net magnetization to the crystal.

To investigate this possibility, we surveyed all magnetic subgroups of the crystallographic 4/*m* point group of $Sr_2IrO_4$ that do not include the two-fold rotation axis (2) as a group element. We find that our data are uniquely but equally well described by two subgroups ($2'/m$ and $m1'$) of the antiferromagnetic point group. Both $2'/m$ and $m1'$ break the global inversion symmetry of the crystal and thus allow a bulk ED contribution to SHG on top of the existing EQ contribution, which is consistent with the large changes observed in the SHG amplitude. Moreover, expressions for the RA-SHG pattern calculated using a coherent sum of the crystallographic EQ and the hidden order induced ED contributions (S2), with the elements of $\chi_{ijkl}^{EQ}$ derived from a $4/m$ point group and those of $\chi_{ijk}^{ED}$ derived from either a $2'/m$ or $m1'$ point group, fit the data extremely well (Fig. 1d). However this indicates that an ordering of higher rank parity-even magnetic multipoles, which preserves global inversion symmetry, cannot be the origin of our RA-SHG results.

On the other hand, our results can be explained by an ordering of higher rank parity-odd magnetic multipoles that preserves the translational symmetry of either the antiferromagnetic structure or the crystallographic structure. A microscopic model that satisfies this condition as well as the $2'/m$ or $m1'$ point group symmetries is the magneto-electric loop-current order (S4), which is predicted to exist in the pseudogap region of the cuprates[14–18,20] but can in principle persist even at half-filling[24]. This phase consists of a pair of counter-circulating current loops in each $CuO_2$ square plaquette (Fig. 1a inset) and can be described by a toroidal pseudovector[25,26] order parameter defined as $\mathbf{\Omega} = \Sigma\, \mathbf{r}_i \times \mathbf{m}_i$, where $\mathbf{r}_i$ is the location of the orbital magnetic moment $\mathbf{m}_i$ inside the plaquette. Four degenerate configurations are possible because the two intra-unit cell current loops can lie along either of two diagonals in the square plaquette and can have either of two time-reversed configurations, which correspond to four 90° rotated directions of the pseudovector. In a real material, one therefore expects domains of all four types to be populated.

To search for domains of the hidden order in $Sr_2IrO_4$ we performed wide-field reflection SHG microscopy measurements. A room temperature $P_{in}$-$S_{out}$ SHG image collected on a clean ~500 µm × 500 µm region parallel to the $ab$ plane is shown in Fig. 2a. This region produces a uniform SHG response consistent with the behaviour of a single crystallographic domain. Upon cooling to $T = 175$ K, brighter and darker patches separated by boundaries that are straight over a tens of micron length scale become visible (Fig. 2b). The distribution and shapes of these patches can be rearranged upon thermal cycling (S5), which suggests that they are not pinned to structural defects in the crystal. To examine whether the different patches observed in Fig. 2b correspond to domains with different $\Omega$ orientation, we performed local RA-SHG measurements within each of the patches. An exhaustive study over the entire crystal area in Fig. 2b reveals a total of only four types of patches characterized by the four distinct RA-SHG patterns (S6) shown in Fig. 3, which are exactly 0°, 90°, 180° and 270° rotated copies of those displayed in Fig. 1. These results are consistent with the hidden phase in $Sr_2IrO_4$ being a magneto-electric loop-current order, but other microscopic models that obey the same set of symmetries certainly cannot be ruled out.

The ordering temperature $T_\Omega$ of the hidden phase in $Sr_2IrO_4$ was determined by measuring the temperature dependence of the SHG intensity in the $P_{in}$-$S_{out}$ polarization geometry where the hidden order induced changes are largest. The onset of change is clearly observed at $T_\Omega \sim 232$ K as shown in Fig. 4(a), which is close to but slightly higher than the value of $T_N \sim 230$ K determined by dc magnetic susceptibility measurements, and evolves with an order parameter-like behaviour upon further cooling (S7). In a Landau free energy expansion (S8), a bilinear coupling between the antiferromagnetic Néel and hidden order parameters is forbidden by symmetry and so there is no a priori reason that $T_\Omega$ and $T_N$ should coincide. The close proximity of $T_\Omega$ and $T_N$ observed in $Sr_2IrO_4$ therefore suggests some microscopic mechanism by which one order can induce the other through a biquadratic coupling, an enhancement of the super-exchange coupling between $J_{eff} = 1/2$ moments due to the presence of hidden order being one possible scenario.

To experimentally examine whether or not the Néel and hidden orders are trivially tied, we performed analogous SHG experiments on hole-doped $Sr_2Ir_{1-x}Rh_xO_4$ crystals (S9) to track the evolution of $T_\Omega$ as a function of Rh concentration ($x$). Bulk magnetization[27] and resonant x-ray diffraction[28] studies have shown that Néel ordering persists for $x \lesssim 0.17$ and that $T_N$ is monotonically suppressed with $x$. But no evidence of any broken symmetry phases beyond the Néel phase has been reported in $Sr_2Ir_{1-x}Rh_xO_4$ to date. Remarkably, our SHG measurements show that the hidden phase transition observed in the parent compound also persists upon hole-doping [Fig. 4(b)-(c)] and that while $T_\Omega$ is suppressed with $x$, the splitting between $T_\Omega$ and $T_N$ grows monotonically from approximately 2 K to 75 K between $x = 0$ and $x \sim 0.11$. This provides strong evidence that the Néel and hidden orders are not trivially tied and that they are independent and distinct electronic phases.

Our finding of a hidden broken symmetry phase in proximity to an antiferromagnetic Mott insulator reveals a striking parallel between the cuprate and $Sr_2IrO_4$ phase diagrams [Fig. 4(d)], which is further strengthened by recent observations of a pseudogap region in $Sr_2Ir_{1-x}Rh_xO_4$ using angle-resolved photoemission spectroscopy[8]. Driving the hidden phase to a quantum critical point[27] through higher doping may be a route to achieving high-$T_c$[12] or parity-odd[29,30] superconductivity in the iridates. While further theoretical studies are required to establish the microscopic origin of the hidden phase, the fact that it bears the symmetries of a magneto-electric loop-current order already suggests several interesting macroscopic responses including a linear magneto-electric effect[18] and non-reciprocal optical rotation[17,18].

**Acknowledgements**

We thank S. Lovesey and D. Khalyavin for providing information about the magnetic point group of the Néel order in $Sr_2IrO_4$. We acknowledge useful discussions with P. Armitage, L. Fu, A. Kaminski, P. A. Lee, O. Motrunich, J. Orenstein, N. Perkins, S. Todadri, C. Varma and V. Yakovenko. This work was support by ARO Grant W911NF-13-1-0059. Instrumentation for the SHG measurements was partially supported by ARO DURIP Award W911NF-13-1-0293. D.H. acknowledges funding provided by the Institute for Quantum Information and Matter, an NSF Physics Frontiers Center (PHY-1125565) with support of the Gordon and Betty Moore Foundation through Grant GBMF1250. R.F. acknowledges the hospitality of the Aspen Center for Physics, supported by NSF Grant PHYS-1066293, where some of this work was carried out. G.C. acknowledges NSF support via Grant DMR-1265162.

Correspondence and requests for materials should be addressed to D.H. (dhsieh@caltech.edu).

**Figure 1**

**Symmetry of the hidden order in $Sr_2IrO_4$.** (a) Crystal structure of a single perovskite layer in $Sr_2IrO_4$ (tetragonal point group $4/m$). Inset shows the basic $IrO_6$ unit of the magneto-electric loop-current order with the arrows pointing in the direction of current flow. (b) Schematic of the RA-SHG experiment. The electric field polarizations of the obliquely incident fundamental beam (in) and outgoing SHG beam (out) can be independently selected to lie either parallel (P) or perpendicular (S) to the light scattering plane (shaded). RA-SHG data are acquired by measuring the SHG intensity $I(2\omega)$ reflected from the (001) surface of $Sr_2IrO_4$ as a function of the angle $\varphi$ between the scattering plane and crystal $ac$-plane. (c) RA-SHG data collected under $P_{in}$-$P_{out}$ and $P_{in}$-$S_{out}$ polarization geometries using $\lambda$ = 800 nm incident light at $T$ = 295 K. (d) Analogous data to panel (c) collected at $T$ = 170 K. All datasets are plotted on the same intensity scale normalized to a value of 1, which corresponds to ~ 20 fW. The high temperature data are fitted to time-reversal invariant bulk electric-quadrupole induced SHG (orange lines). The low temperature data can only be fitted to the coherent sum of time-reversal invariant bulk electric-quadrupole and time-reversal broken bulk electric-dipole induced SHG (purple lines) as described in the text.

**Figure 2**

**Spatial mapping of hidden order domains.** (a) Wide-field reflection SHG microscopy image of the cleaved (001) plane of $Sr_2IrO_4$ measured under $P_{in}$-$S_{out}$ polarization geometry at $\varphi$ = 78° at $T$ = 295 K and (b) at $T$ = 175 K. A patchwork of light and dark regions present in the low temperature image arises from domains of the hidden order. The dark curves at the sample edge and dark micron-sized spots near the sample center that are visible in both the high and low temperature images are structural defects.

**Figure 3**

**Degenerate ground-state configurations of the hidden order.** Four different types of RA-SHG patterns found by performing local measurements within all of the domains mapped in Fig. 2b. Red lines are fits to the expressions described in the text. The large lobes are shaded pink to emphasize the orientation of each pattern. Schematic of the four degenerate magneto-electric loop-current order configurations are shown below each pattern to illustrate the possible correspondence. The red arrows denote the direction of the toroidal moment $\Omega$ in each plaquette

**Figure 4**

**Temperature and hole-doping dependence of the hidden order.** (a) The change in SHG intensity from $Sr_2Ir_{1-x}Rh_xO_4$ at $x$ = 0, (b) $x$ = 0.04 and (c) $x$ = 0.11 measured relative to their room temperature values as a function of temperature. Data were taken under $P_{in}$-$S_{out}$ polarization geometry at $\varphi$ = 78°. No difference was observed between curves measured upon cooling and heating. Lines are guides to the eye. The dashed red lines mark the transition temperature $T_\Omega$ of the hidden order phase deduced from our SHG data and the dashed black lines mark the Néel temperature $T_N$ determined from dc magnetic susceptibility measurements. (d) Temperature versus doping phase diagram of $Sr_2Ir_{1-x}Rh_xO_4$ showing the boundaries of the hidden order and the long-range (LRO) and short-range (SRO) Néel ordered regions. Points where a pseudogap is present are also marked, although a pseudogap phase boundary is not yet experimentally known.

**Methods**

**Material growth:** Single crystals of $Sr_2IrO_4$ and $Sr_2Ir_{1-x}Rh_xO_4$ were grown using a self flux technique from off-stoichiometric quantities of $IrO_2$, $SrCO_3$ and $SrCl_2$ or $RhO_2$, $IrO_2$, $SrCO_3$ and $SrCl_2$ respectively. The ground mixtures of powders were melted at $1470°C$ in partially capped platinum crucibles. The soaking phase of the synthesis lasted for >20 hours and was followed by a slow cooling at $2°C/hr$ to reach $1400°C$. From this point the crucible is brought to room temperature through a rapid cooling at a rate of $100°C/hr$. The Rh concentration was determined by energy-dispersive x-ray spectroscopy.

**RA-SHG measurements:** Ultrashort optical pulses with 35 fs duration and 800 nm center wavelength were produced at a 10 kHz repetition rate from a regeneratively amplified Ti:sapphire laser (KMLabs - Wyvern). The RA-SHG measurements were performed using a rotating scattering plane technique[31] utilizing an SHG wavelength $\lambda = 400$ nm that is resonant with the O 2p to $J_{eff} = 1/2$ upper Hubbard band transition[32]. Light was obliquely incident onto the sample at a 30° angle of incidence with a fluence that was maintained below 1 mJ/cm$^2$, which is well below the damage threshold of $Sr_2IrO_4$. Reflected SHG light was collected using a photomultiplier tube. Samples were cleaved either in air or in a nitrogen purged environment and immediately pumped down to a pressure $< 5 \times 10^{-6}$ torr in an optical cryostat. Crystals were oriented prior to measurement using x-ray Laue diffraction.

**SHG microscopy measurements:** Wide field SHG images with a spatial resolution of ~ 1 µm were collected using the same light source as that used for the RA-SHG measurements. A 40° oblique angle of incidence was used and the incident light fluence was maintained below 1 mJ/cm$^2$. The reflected SHG light was collected by an objective lens and focused onto an electron-multiplying CCD camera.

**Fitting procedure:** The low ($T < T_\Omega$) and high temperature ($T > T_\Omega$) RA-SHG data were fit to the expressions
$$I(2\omega,\varphi) = \left| A\hat{e}_i(2\omega)\chi_{ijkl}^{EQ}(\varphi)\hat{e}_j(\omega)\partial_k\hat{e}_l(\omega) + A\hat{e}_i(2\omega)\chi_{ijk}^{ED}(\varphi)\hat{e}_j(\omega)\hat{e}_k(\omega) \right|^2 I(\omega)^2 \text{ and}$$

$I(2\omega,\varphi) = \left| A\hat{e}_i(2\omega)\chi^{EQ}_{ijkl}(\varphi)\hat{e}_j(\omega)\partial_k\hat{e}_l(\omega) \right|^2 I(\omega)^2$ respectively, where *A* is a constant determined by the experimental geometry, $I(\omega)$ is the intensity of the incident beam, $\hat{e}$ is the polarization of the incoming fundamental or outgoing SHG light and $\chi^{EQ}_{ijkl}(\varphi)$ and $\chi^{ED}_{ijk}(\varphi)$ are the bulk electric-quadrupole and electric-dipole susceptibility tensors respectively transformed into the rotated frame of the scattering plane. The non-zero independent elements of the tensors in the un-rotated frame of the crystal are deduced by applying the appropriate point group ($4/m$ for $\chi^{EQ}_{ijkl}$ and $2'/m$ or $m1'$ for $\chi^{ED}_{ijk}$) and degenerate SHG permutation symmetries. This reduces $\chi^{EQ}_{ijkl}$ to 17 non-zero independent elements (*xxxx* = *yyyy*, *zzzz*, *zzxx* = *zzyy* = *zxxz* = *zyyz*, *xyzz* = −*yxzz*, *xxyy* = *yyxx* = *xyyx* = *yxxy*, *xxxy* = −*yyyx* = *xyxx* = −*yxyy*, *xxzz* = *yyzz*, *zzxy* = −*zzyx* = −*zxyz* = *zyxz*, *xyxy* = *yxyx*, *xxyx* = −*yyxy*, *zxzx* = *zyzy*, *xzyz* = −*yzxz*, *xzxz* = *yzyz*, *zxzy* = −*zyzx*, *yxxx* = −*xyyy*, *xzzx* = *yzzy*, *xzzy* = −*yzzx*) and reduces $\chi^{ED}_{ijk}$ to 10 non-zero independent elements (*xxx*, *xyx* = *xxy*, *xyy*, *xzz*, *yxx*, *yyx* = *yxy*, *yyy*, *yzz*, *zzx* = *zxz*, *zzy* = *zyz*).

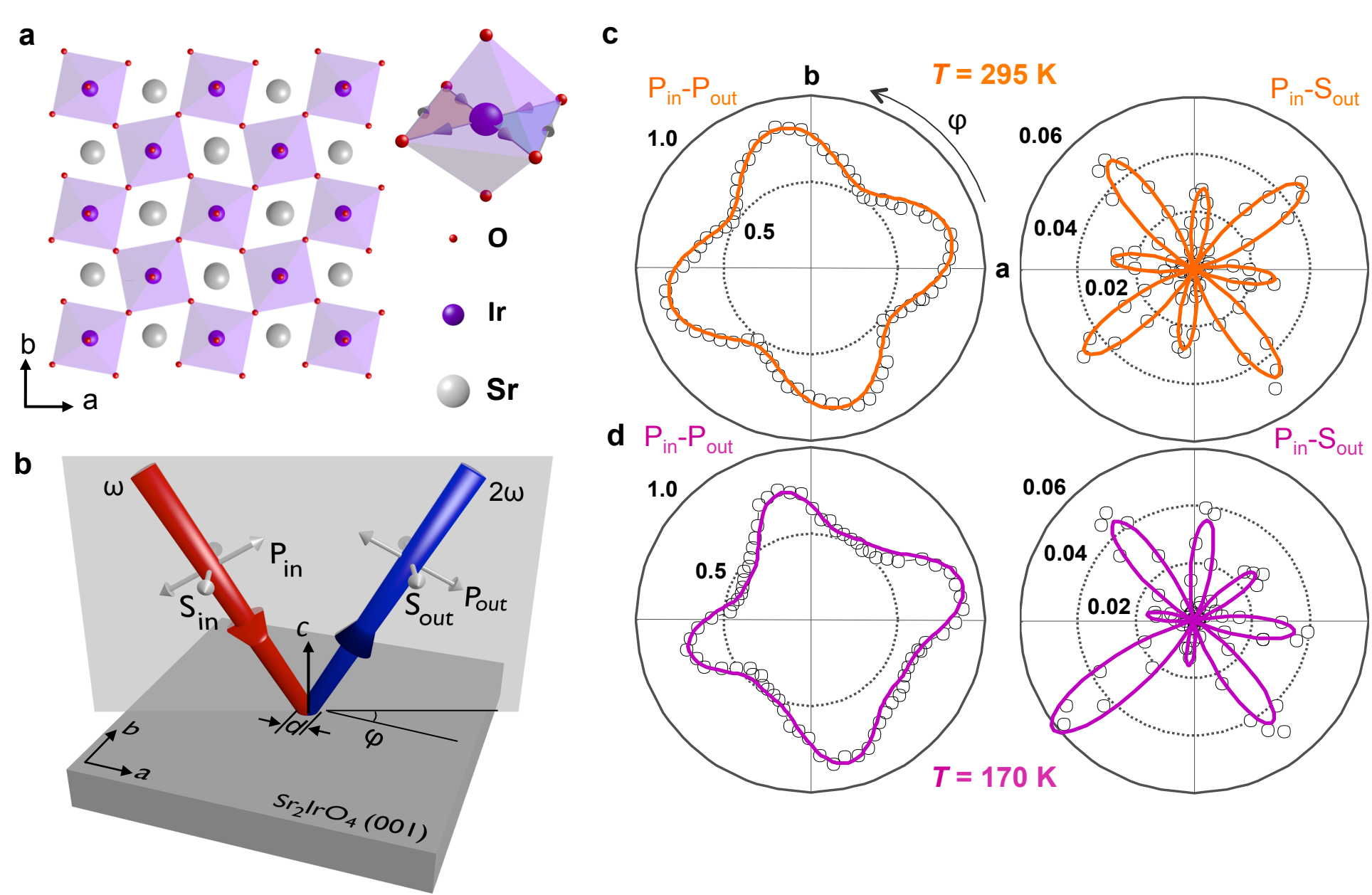

Figure 1

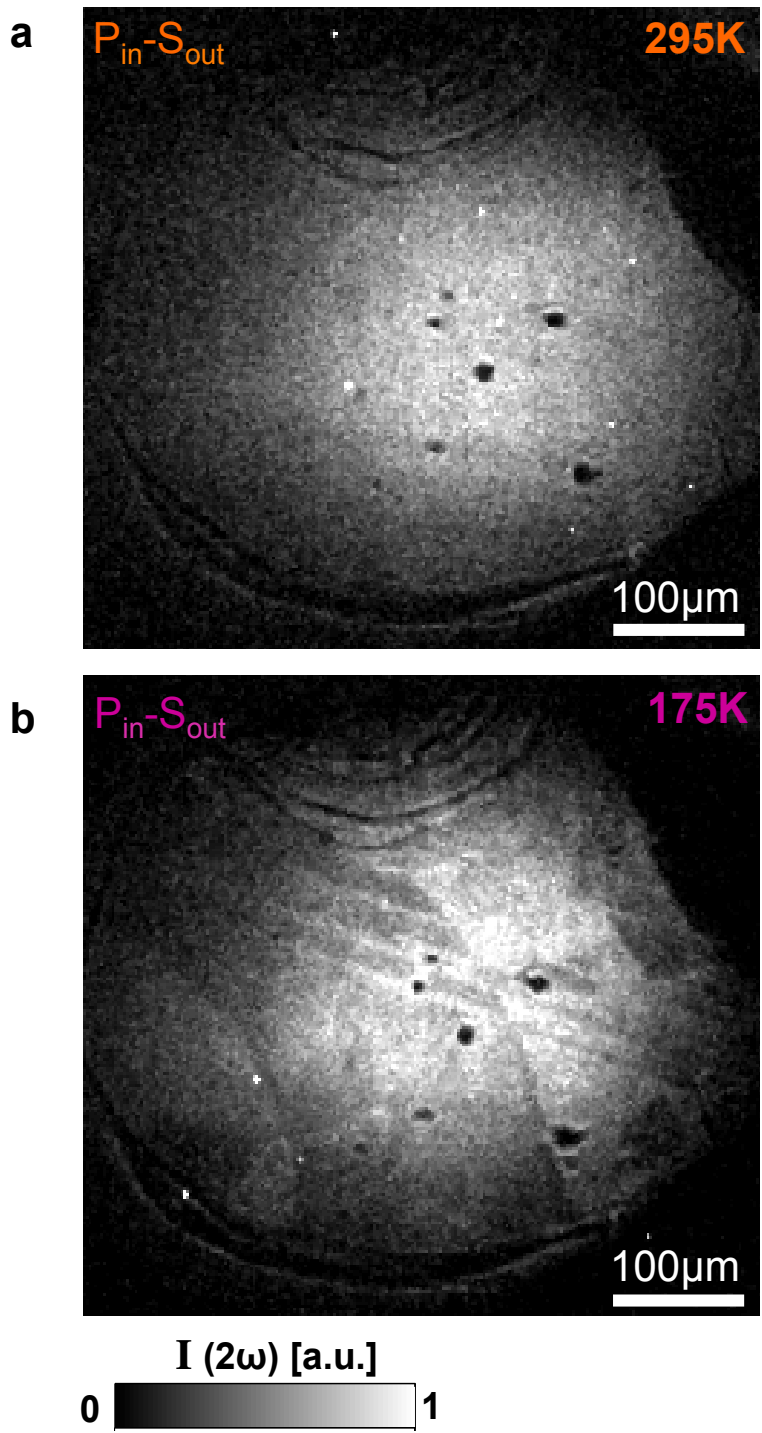

Figure 2

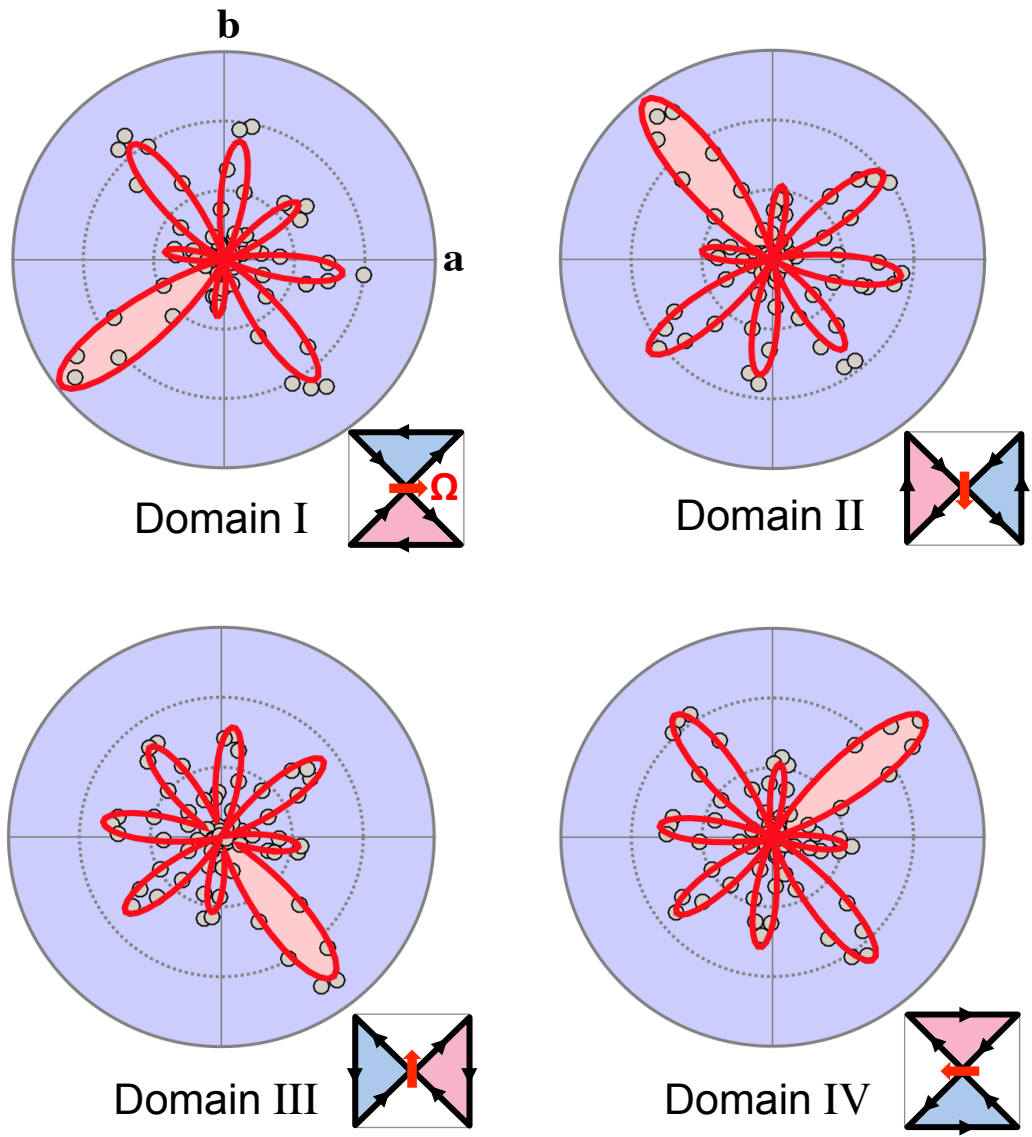

Figure 3

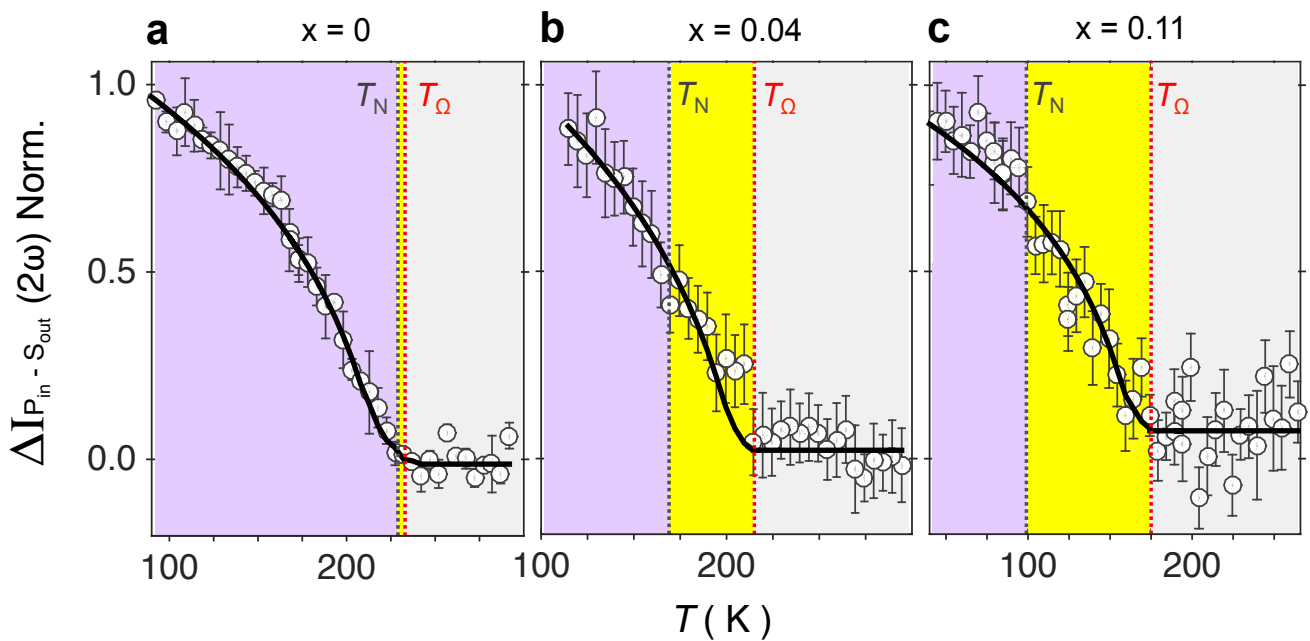
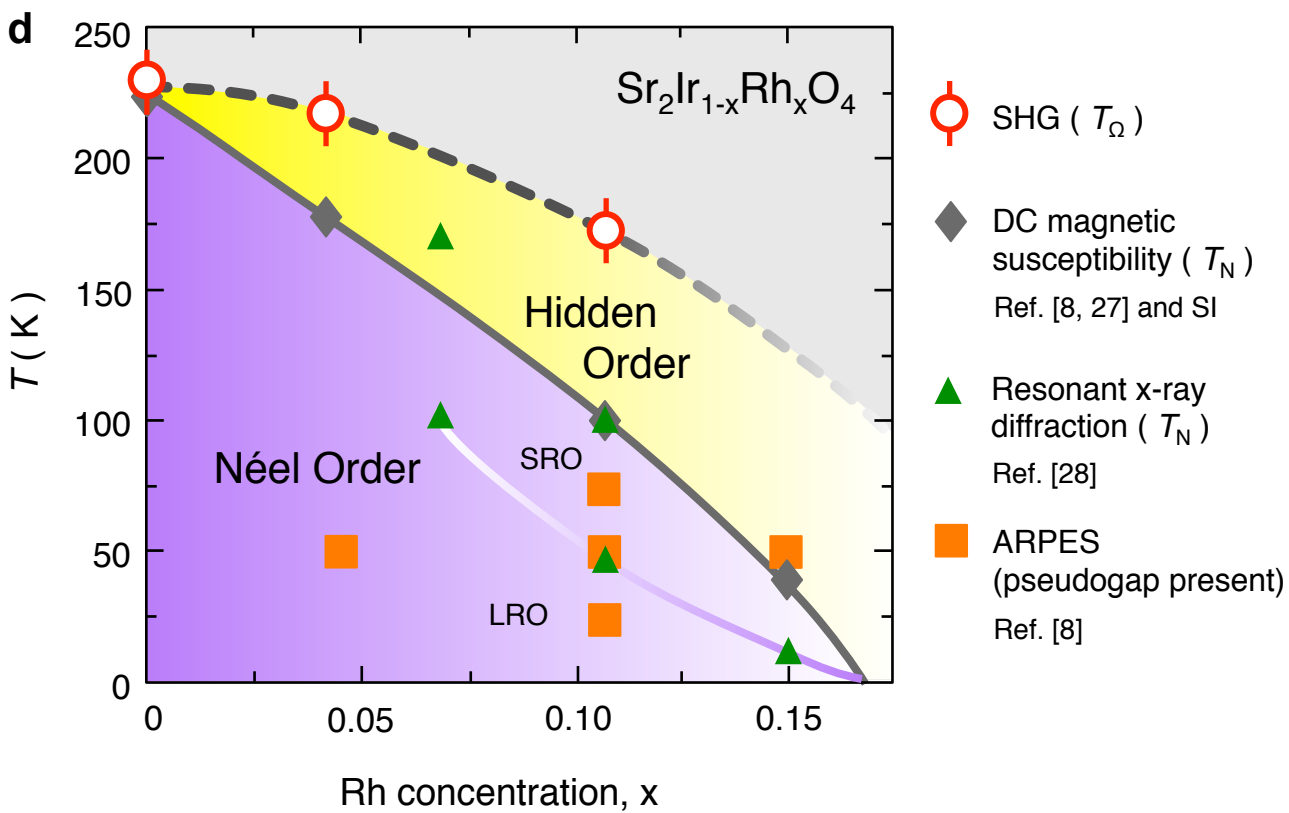

Figure 4

Supplementary Information for

**Evidence of an odd-parity hidden order in a spin-orbit coupled correlated iridate**

Contents:

# S1. RA-SHG data for $S_{in}$-$P_{out}$ and $S_{in}$-$S_{out}$ geometries above and below $T_\Omega$

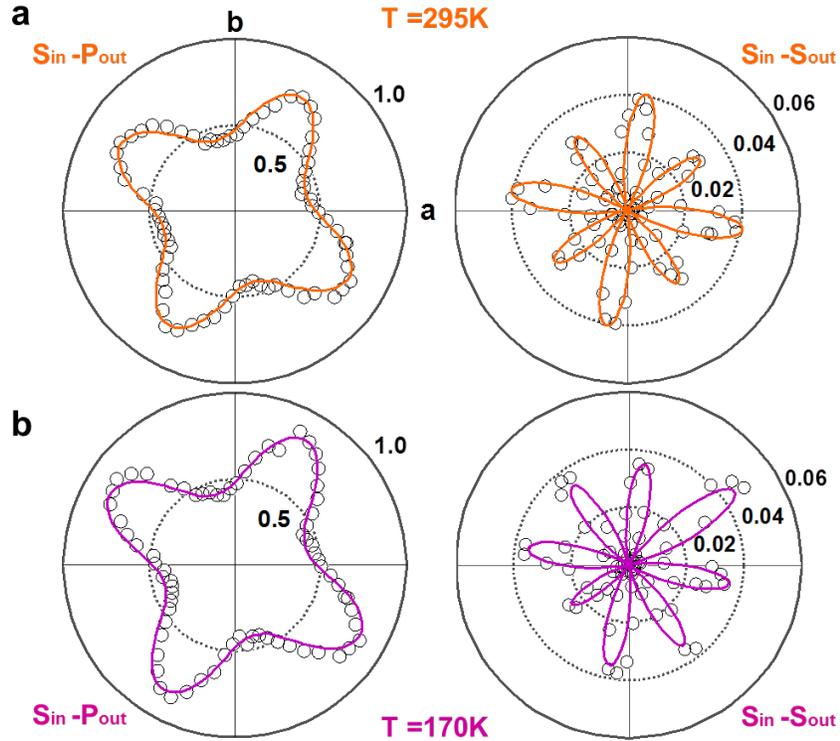

Fig. S1. RA-SHG data for $S_{in}$-$P_{out}$ (left column) and $S_{in}$-$S_{out}$ (right column) polarization geometries collected at (a) $T = 295$ K and (b) $T = 170$ K. The tetragonal crystallographic axes *a* and *b* are labeled in the upper left plot. Lines are best fits to expressions given in Section S2.

RA-SHG data collected under two other polarization geometries $S_{in}$-$P_{out}$ and $S_{in}$-$S_{out}$, which are complementary to those shown in Figs 1c & d of the main text, are shown for completeness in Fig. S1. The lowering of rotational symmetry from $C_4$ to $C_1$ upon cooling below $T_\Omega$ is visible in both $S_{in}$-$P_{out}$ and $S_{in}$-$S_{out}$ geometries. The effect is present but less pronounced in $S_{in}$-$P_{out}$ geometry owing to the different set of susceptibility tensor elements probed (Section S2). The high temperature data in Fig. S1 are very well described by bulk EQ-induced SHG from a crystallographic 4/*m* point group while the low temperature data are very well described by a coherent sum of an EQ contribution (4/*m* point group) and an ED contribution (2′/*m* or *m*1′ magnetic point group). These results are fully consistent with Figs 1c & d of the main text.

## S2. Mathematical expressions used for fitting RA-SHG patterns

i) Fitting RA-SHG data for $T > T_\Omega$

Recent neutron diffraction[1,2], RA-SHG and optical third harmonic generation measurements[3] show that the crystal structure of $Sr_2IrO_4$ belongs to the centrosymmetric tetragonal $4/m$ point group as opposed to the previously accepted $4/mmm$ point group[4,5]. Given the presence of inversion symmetry, the leading order contribution to SHG is the non-local term of electric-quadrupole type, which can be expressed as an effective nonlinear polarization as $P_i(2\omega) \propto \chi^{EQ}_{ijkl} E_j(\omega) \nabla_k E_l(\omega)$ where $\chi^{EQ}_{ijkl}$ is the electric-quadrupole susceptibility tensor. By enforcing $4/m$ point group symmetry, $\chi^{EQ}_{ijkl}$ is reduced to having 21 non-zero independent elements[6]:

$$\{xxxx = yyyy, \; zzzz, \; zzxx = zzyy, \; xyzz = -yxzz, \; xxyy = yyxx,$$
$$xxxy = -yyyx, \; xxzz = yyzz, \; zzxy = -zzyx, \; xyxy = yxyx, xxyx = -yyxy,$$
$$zxzx = zyzy, \; xzyz = -yzxz, \; xyyx = yxxy, \; xyxx = -yxyy, \; xzxz = yzyz,$$
$$zxzy = -zyzx, \; yxxx = -xyyy, \; zxxz = zyyz, \; zxyz = -zyxz, \; xzzx = yzzy,$$
$$xzzy = -yzzx\}$$

With the four additional constraints from degenerate SHG $\{zzxx = zxxz, zzyx = zxyz, xxyy = xyyx, xxxy = xyxx\}$, the number of non-zero independent tensor elements is further reduced to 17. The rotation of the crystal by an angle φ about the $c$-axis is carried out mathematically by applying a basis transformation on the reduced tensor from the original (primed) to rotated (un-primed) reference frame using

$$\chi^{EQ}_{ijkl}(\varphi) = R_{ii'}(\varphi) R_{jj'}(\varphi) R_{kk'}(\varphi) R_{ll'}(\varphi) \chi^{EQ}_{i'j'k'l'}$$

where $R_{ij}(\varphi)$ is the rotation matrix about the $c$-axis. Finally, the expression that is used to fit the RA-SHG data at $T > T_\Omega$ is given by $I(2\omega, \varphi) = \left| A \hat{e}_i(2\omega) \chi^{EQ}_{ijkl}(\varphi) \hat{e}_j(\omega) \partial_k \hat{e}_l(\omega) \right|^2 I(\omega)^2$, where $A$ is a constant determined by the experimental geometry, $I(\omega)$ is the intensity of the incident

beam and $\hat{e}$ is the polarization of the incoming or outgoing light, which we select to be either linearly P or S polarized. We note that previous work has already shown that there is no evidence of a surface electric-dipole contribution SHG[3] and that the crystallographic symmetry of the surface remains unchanged across $T_\Omega$[7].

ii)  Fitting RA-SHG data for $T < T_\Omega$

The low temperature RA-SHG data are fit to a coherent sum of the electric-quadrupole term described above and a hidden order induced electric-dipole term. The electric-dipole contribution is expressed as a nonlinear polarization $P_i(2\omega) \propto \chi^{ED}_{ijk} E_j(\omega) E_k(\omega)$. By enforcing $2'/m$ magnetic point group symmetry, $\chi^{ED}_{ijk}$ is reduced to having 14 non-zero independent elements:

$$\{xxx, xyx, xxy, xyy, xzz, yxx, yyx, yxy, yyy, yzz, zzx, zzy, zxz, zyz\}$$

We only discuss the results using a $2'/m$ magnetic point group here although the same procedure was applied to all of the magnetic point groups we surveyed. The additional constraints from degenerate SHG $\{xyx = xxy, yyx = yxy, zzx = zxz, zzy = zyz\}$ leaves 10 non-zero independent tensor elements remaining. A basis transformation was then carried out on $\chi^{ED}_{ijk}$ using $\chi^{ED}_{ijk}(\varphi) = R_{ii'} R_{jj'} R_{kk'} \chi^{ED}_{i'j'k'}$ and the expression used to fit the RA-SHG data at $T < T_\Omega$ is

$$\tilde{I}(2\omega, \varphi) = \left| A\hat{e}_i(2\omega) \chi^{EQ}_{ijkl}(\varphi) \hat{e}_j(\omega) \partial_k \hat{e}_l(\omega) + A\hat{e}_i(2\omega) \chi^{ED}_{ijk}(\varphi) \hat{e}_j(\omega) \hat{e}_k(\omega) \right|^2 I(\omega)^2.$$

## S3. Visualizing the 2/$m$1′ magnetic point group symmetry of the Néel phase

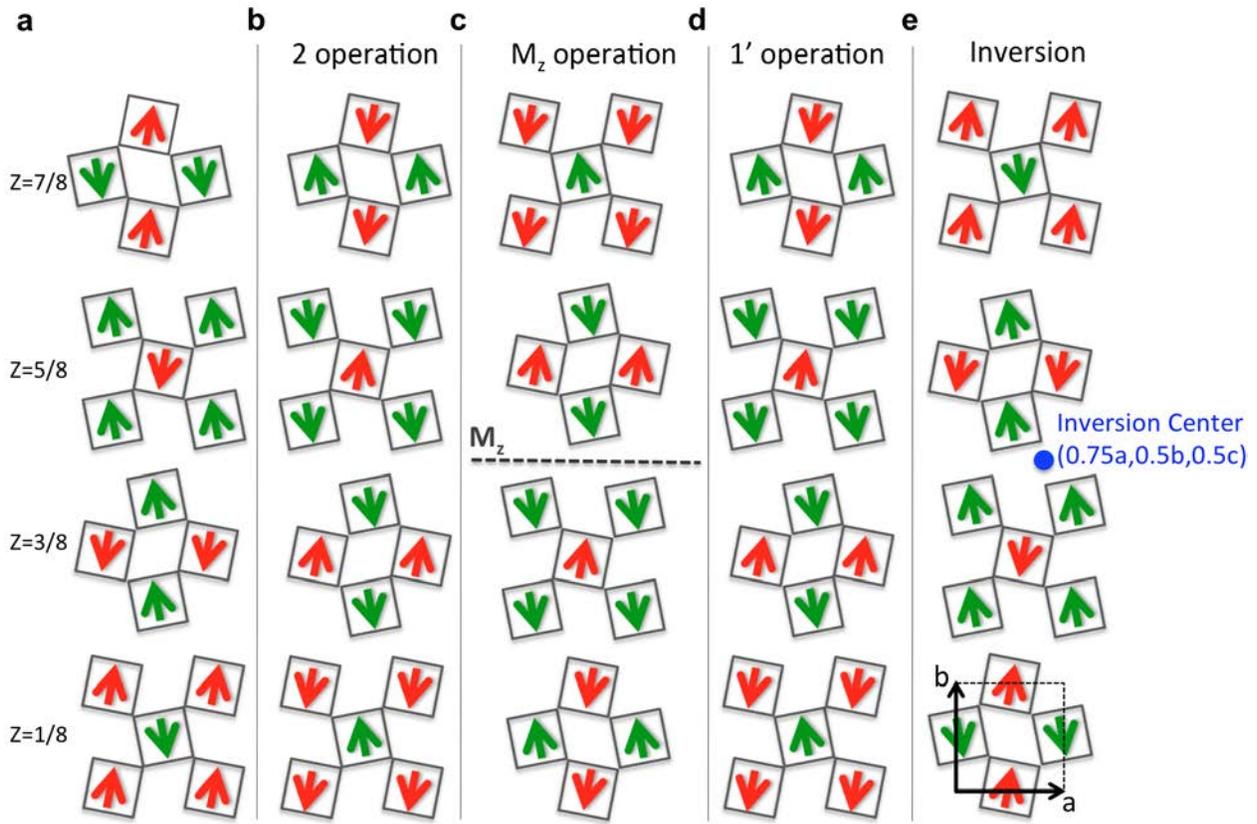

Fig. S2 (a) Known dipolar antiferromagnetic structure of $Sr_2IrO_4$. Planes through each Ir-O layer in the unit cell are shown, with z denoting the position of the layer along the *c*-axis. The red and green arrows denote the direction of the magnetic dipole moments in the two structural sub-lattices. Resultant magnetic structure upon applying the following operations contained within the 2/$m$1′ point group: (b) 180° rotation about the *c*-axis, (c) reflection about a mirror plane normal to the *c*-axis, (d) time-reversal, (e) spatial inversion. All structures are related by simple lattice translation.

The antiferromagnetic structure of $Sr_2IrO_4$ reported by neutron and resonant x-ray diffraction measurements is shown in Fig. S2a. Its magnetic point group derived from the tetragonal 4/$m$ crystallographic point group of $Sr_2IrO_4$ is 2/$m$1′ [Courtesy of S. Lovesey and D. Khalyavin]. Invariance of the magnetic structure under the elements of 2/$m$1′ is explicitly shown in Figs S2 b-e. We note that the 2/$m$1′ magnetic point group assignment does not rely on the magnitude of the magnetic moments on the two structural sub-lattices being equal, even though experimentally they are found to be so.

## S4. Loop-current order in cuprates versus iridates

Loop-current ordered phases were initially proposed in a three-band $CuO_2$ model of the copper-oxide planes where the charge currents are emergent complex hopping terms between oxygen and copper sites[8,9]. In this section, we examine how this model would be modified for $Sr_2IrO_4$. Our intention is not to elaborate on any of the details of loop-current order – for that, the reader should see the references[8,9] – but rather to elucidate the microscopically relevant terms and how the model will differ between the cuprates and the iridates if it were indeed present in $Sr_2IrO_4$.

Loop-current order in the iridates should be captured within a modified Emery model[10] of the $IrO_2$ plane that treats the oxygen p-holes ($p_{i\sigma}^\dagger$) and iridum $t_{2g}$ d-holes ($d_{j\alpha\sigma}^\dagger$), where $\sigma$ is the spin index, and $\alpha$ labels the three $t_{2g}$ bands:

$$H = H_{Ir} + H_O + t_{pd} \sum_{\langle ij \rangle} \sum_{\alpha\sigma} (p_{i\sigma}^\dagger d_{j\alpha\sigma} + H.c.) + V_{pd} \sum_{\langle ij \rangle} n_{pi} n_{dj}.$$

$H_{Ir}$ captures the $J_{eff} = 1/2$ and $J_{eff} = 3/2$ bands as well as any direct Ir-Ir hopping $t_{dd}$. All the effects of the strong spin-orbit coupling as well as the Hubbard $U_d$ are contained within $H_{Ir}$. This part of the model has been discussed extensively elsewhere[11,12] and will not be repeated here. While the $J_{eff}$ language is more appropriate to describe the low energy Ir on-site excitations, it is single spin-½ electrons that hop between the iridium and oxygen sites via $t_{pd}$ and give rise to loop-current order, so here we use this language. In addition to the iridium-oxygen hopping, there is a nearest-neighbor Coulomb interaction $V_{pd}$ between Ir and O sites, where $n_{pi}$ and $n_{dj}$ describe the total density of p-electrons and d-electrons at particular sites. The term $H_O$ describes the oxygen bands:

$$H_O = -t_{pp} \sum_{\langle ij \rangle} p_{i\sigma}^\dagger p_{j\sigma} + H.c. + V_{pp} \sum_{\langle ij \rangle} n_{p,i} n_{p,j} + \mu_p \sum_i n_{p,i} + U_p \sum_i n_{p,i\uparrow} n_{p,i\downarrow}$$

which includes a chemical potential $\mu_p$, on-site Hubbard interaction $U_p$, hopping $t_{pp}$ and a nearest-neighbor Coulomb interaction $V_{pp}$.

Loop-current order requires these nearest-neighbor Coulomb interaction terms ($V_{pd}$ and $V_{pp}$), as it emerges from a mean-field decoupling of these terms that generates complex contributions to the iridium-oxygen and oxygen-oxygen hopping,

$$\tilde{t}_{pd}^{ij} \propto V_{pd} \langle p_{i\sigma}^{\dagger} d_{j\alpha\sigma} \rangle; \quad \tilde{t}_{pp}^{ij} \propto V_{pp} \langle p_{i\sigma}^{\dagger} p_{j\sigma} \rangle \ .$$

As it is a Fock decoupling, like spin density waves it is favored by larger interactions, $V_{pd}$ and $V_{pp}$, with a critical threshold for order. The deeper in energy the oxygen states are, the larger these critical thresholds will be, however hole-doping is not required to have loop-current order. In Varma's original theory, $t_{pd}$ was assumed to grow linearly with the hole-doping, $x$, which suppressed loop-current order near half-filling[9]. However if this phenomenological assumption is removed, it has been theoretically shown that loop-current order can exist at half-filling[13]. Of course, whether loop-current order beats out or coexists with other possible orders is highly dependent on the microscopic details; here we simply address the stability of loop-current order independent of other orders.

While the iridates mimic the cuprates in many ways, there are several key differences between these two systems that can affect the stability of loop-current order:

• The cuprates are charge-transfer insulators, meaning that any doped holes go into the oxygen states, forming Zhang-Rice singlets, and that the stability of loop-current order is expected to increase with doping[9]. This is the phenomenological justification for assuming $t_{pd} \propto x$ in Varma's original theory. On the other hand, the oxygen bands in the iridates are 2-3 eV below $E_F$ and doped holes are expected to go into the $J_{\text{eff}} = 1/2$ level[14]. Therefore, unlike in the cuprates, loop-current order is not expected to be stabilized by hole-doping in the iridates.

• The iridium 5$d$ states are significantly larger than the copper 3$d$ states, which leads to direct Ir-Ir hopping and exchange terms as well as a significantly larger $t_{pd}$ and $V_{pd}$ that should favor loop-current order in iridates. Concomitantly, the in-plane O-O distance is slightly larger in $Sr_2IrO_4$ (2.81 Å) than, for example, in $HgBa_2CuO_4$, (2.74 Å), although the apical to in-plane O-O distance is significantly smaller (2.87 Å vs. 3.34 Å respectively)[15].

- The Hubbard $U_d$ in the iridates is significantly smaller, and whether $Sr_2IrO_4$ is more a Mott or Slater insulator is still unresolved[7,16,17]. It is currently unclear how the on-site iridium interactions will affect the stability of loop-current order.

- The strong spin-orbit coupling of the iridates should not significantly affect the nature of the loop-current order, as it involves the direct hopping and interactions of single spin-½ electrons from Ir-O and O-O. Of course, it will affect the resulting phase diagram, and details may change, but the general nature should be unaffected.

In conclusion, it would be a bit surprising to find loop-current order in the iridates given the depth of the oxygen bands ($\mu_p$). However, given that spiral magnetic and antiferroelectric orders are not observed experimentally, all the remaining explanations of the toroidal order parameter involve the oxygens somehow, either via loop-current order or by developing oxygen moments. Due to the depth of the oxygen bands, loop-current order is the more likely of the options, and indeed the larger $V_{pd}$ and $V_{pp}$ will likely favor this order. The key difference from the cuprates is that hole-doping is not required to drive loop-current order in the iridates, and so it can exist at half-filling.

One of the simplest models of loop-current order proposed in the cuprates is the magneto-electric loop-current ordered phase otherwise known as the $\Theta_{II}$ loop-current ordered phase. Its magnetic point group[18], which is derived from the tetragonal 4/*mmm* crystallographic point group of the cuprate lattice, is *m'mm*. Because the $Sr_2IrO_4$ lattice belongs to a lower symmetry subgroup (4/*m*) of the cuprate lattice (4/*mmm*) owing to the absence of the *ac* and *bc* mirror planes, the $\Theta_{II}$ loop current ordered phase in $Sr_2IrO_4$ also belongs to the corresponding lower symmetry subgroup (2'/*m*) of the cuprate $\Theta_{II}$ loop current ordered phase (*m'mm*) with missing *ac* and *bc* mirror planes as shown explicitly in Fig. S3.

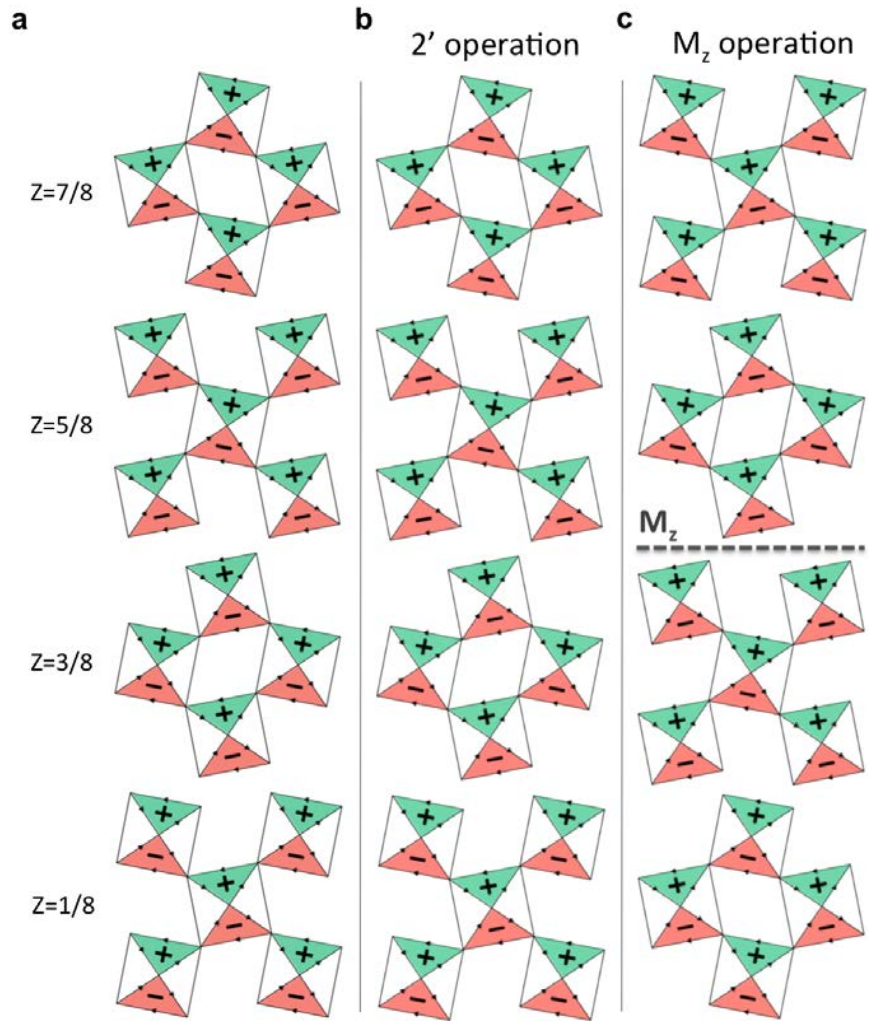

Fig. S3. The $\Theta_{II}$ loop-current ordered phase shown in panel (a) is invariant under the operations contained within the $2'/m$ point group: (b) 180° rotation about the $c$-axis multiplied by time-reversal and (c) reflection about a mirror plane normal to the $c$-axis. All structures are related by simple lattice translation. "−" indicates a clockwise and "+" indicates a counter-clockwise loop-current.

More exotic loop-current configurations can be constructed to satisfy the $m1'$ point group symmetries such as the one illustrated in Fig. S4. Similar to the $\Theta_{II}$ loop-current ordered phase, this configuration breaks global inversion symmetry and two-fold rotational symmetry as shown explicitly in Fig. S4e. Although such an order cannot be ruled out based on our SHG data alone, it has no precedence in the cuprate literature.

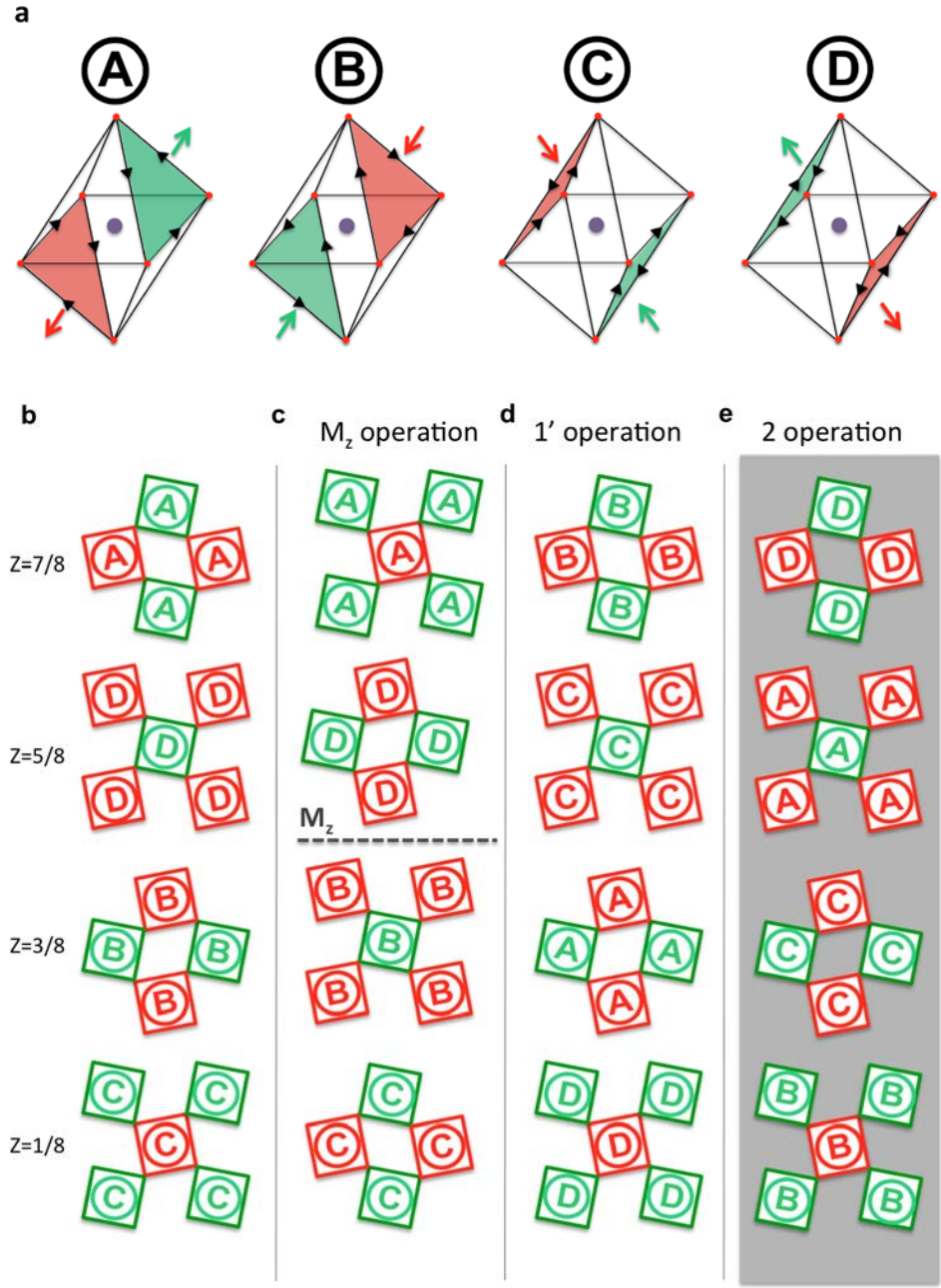

Fig. S4. A loop-current order configuration (b) that satisfies $m1'$ can be constructed using the four octahedral building blocks shown in panel (a), which are labeled A, B, C and D. The four building blocks are related via $\mathbf{M_z}(A) \to C$, $\mathbf{M_z}(B) \to D$, $\mathbf{TR}(A) \to B$, $\mathbf{TR}(C) \to D$, $\mathbf{R_2}(A) \to D$ and $\mathbf{R_2}(B) \to C$, where the operations $\mathbf{M_z}$, $\mathbf{TR}$ and $\mathbf{R_2}$ represent reflection about a mirror plane normal to $z$-axis, time reversal and 180° rotation about $z$-axis respectively. Panels (c) and (d) show that both $\mathbf{M_z}$ and $\mathbf{TR}$ leave the configuration invariant modulo a lattice translation, which is consistent with the $m1'$ point group assignment. Panel (e) explicitly shows that two-fold rotational symmetry is broken because a rotation by 180° alters the stacking sequence from ADBC to DACB, which are not related by a lattice translation.

## S5. Behavior of hidden order domains under repeated thermal cycles

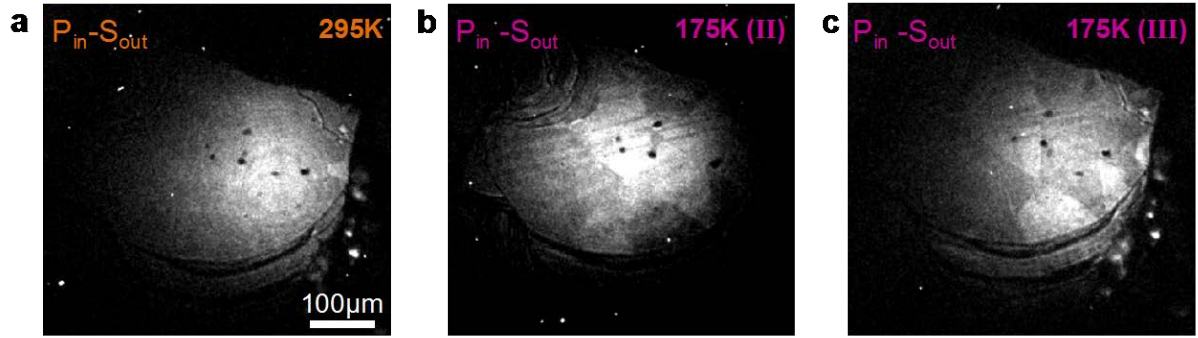

Fig. S5. SHG image taken under P$_{in}$-S$_{out}$ geometry at a temperature (a) above $T_\Omega$ and below $T_\Omega$ after (b) the second and (c) the third cool downs. These should be compared to Fig. 2 of the main text, which was the first cool down.

Fig. S5 shows SHG images analogous to Fig. 2 of the main text taken after consecutive thermal cycles across $T_\Omega$. Fig. S5b and c show images taken after the second and third cool downs, which means that the crystal has undergone one or two thermal cycles respectively after the data in Fig. 2b was taken. It is clear that the domain size and distributions are rearranged following each thermal cycle across $T_\Omega$, which indicates that the hidden ordered domains are not pinned to any structural defects in the material.

## S6.  Survey of RA-SHG patterns of $Sr_2IrO_4$ across spatial regions

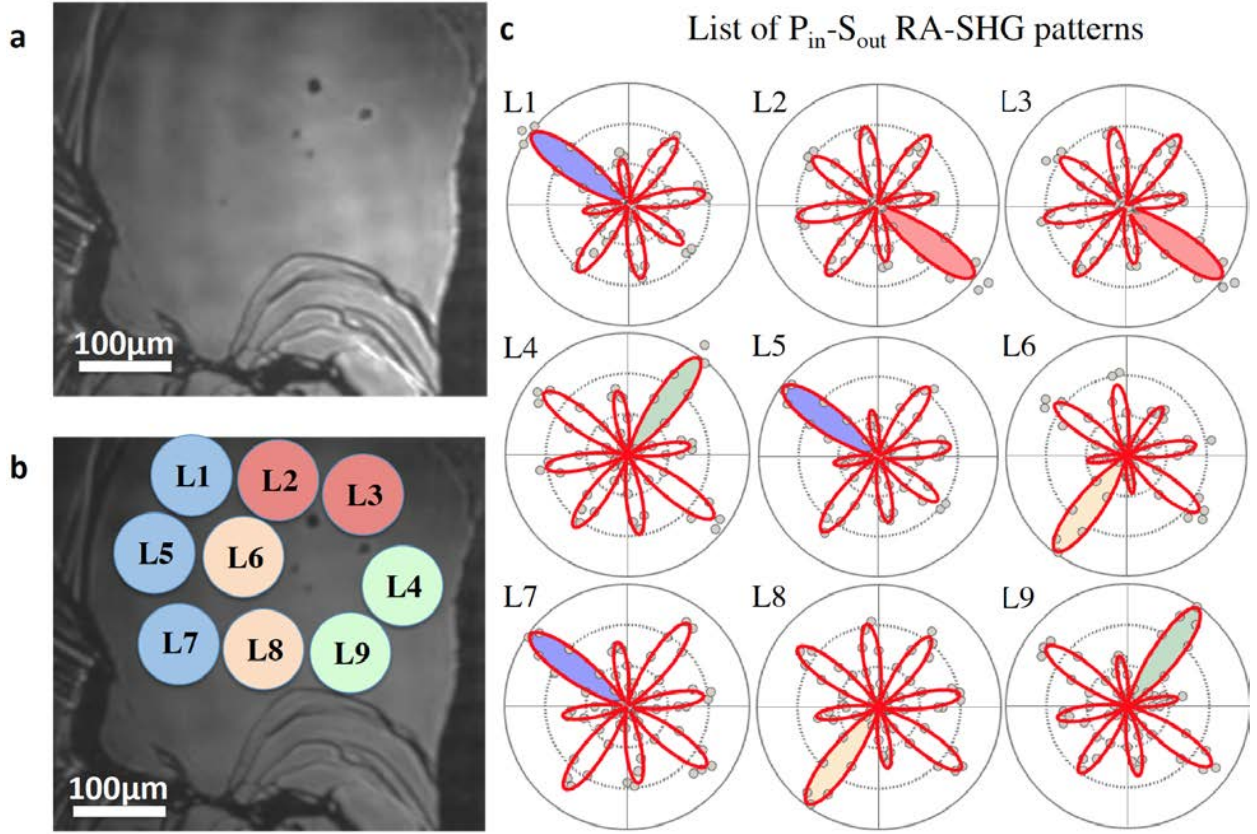

Fig. S6. (a) Conventional optical microscope image of the cleaved surface of a typical $Sr_2IrO_4$ crystal. (b) RA-SHG patterns show the same symmetry across all locations on the surface. A select few locations, denoted L1 through L9, are displayed in this figure. (c) RA-SHG patterns for $P_{in}$-$S_{out}$ polarization geometry collected at $T = 175$ K from points L1 through L9. Locations exhibiting the same domain orientation are marked by the same color.

In the antiferromagnetic phase of $Sr_2IrO_4$, the canting of the $J_{eff} = 1/2$ moments produces a small net in-plane ferromagnetic moment in each layer, which orders in an ↑-↓-↓-↑ antiferromagnetic pattern along the $c$-axis (Fig. S2a). It has been reported that above a critical magnetic field of approximately 0.2 T, the stacking sequence along the $c$-axis changes to a ferromagnetic ↑-↑-↑-↑ pattern[19], which imparts an overall net in-plane magnetization to the crystal and thereby reduces the rotational symmetry of the crystal around the $c$-axis from $C_4$ to $C_1$.

Although our experiments are performed in zero magnetic field, which rules out the possibility of a global ferromagnetic ↑-↑-↑-↑ stacking throughout the crystal, we would also like to rule out the possibility that we are sampling isolated imperfect regions of the crystal where a ↑-↑-↑-↑ stacking is somehow locally realized even in zero magnetic field. To do so, we performed spatially dependent RA-SHG measurements below $T_\Omega$ by scanning the beam location over the entire surface of $Sr_2IrO_4$ crystals. Figure S6a shows an optical image of the cleaved (001) surface of one of the $Sr_2IrO_4$ crystals used in this study. All locations surveyed on the surface consistently show the same symmetry below $T_\Omega$ (Fig. S6b & c) and is reproducible across different crystals, which were obtained from the same batch used for magnetic neutron diffraction measurements published elsewhere[2]. These results show that the symmetry we report for $Sr_2IrO_4$ below $T_\Omega$ is representative of the bulk of the crystals and does not arise from isolated regions with imperfect layer stacking.

We also note that a ↑-↑-↑-↑ stacked magnetic structure does not break inversion symmetry and therefore is anyways not consistent with our assigned 2'/m or m1' point groups. This is further corroborated by the fact that no anomaly is observed below $T_N$ in the 11% Rh doped $Sr_2IrO_4$ crystals (Fig. 4c), which was found by resonant x-ray diffraction to adopt the ↑-↑-↑-↑ stacked magnetic structure[20].

## S7. Functional form of the SHG temperature dependence

Figures 4a-c of the main text show that the temperature dependence of the SHG intensity exhibits an order-parameter like behavior that can be fit to a function of the form $I(2\omega) = a|T - T_\Omega|^\beta$. However, we do not report this value because the fitted exponent cannot be directly related to a critical exponent of the order parameter for the following reason.

The low temperature SHG intensity is a coherent sum of a bulk EQ contribution that is present at all temperatures and a bulk ED contribution that only turns on below $T_\Omega$. The temperature dependence of the latter contribution, embedded in $\chi^{ED}_{ijk}(\varphi, T)$, provides information about the critical behavior of the phase transition. Using the expression derived in Section S2, it is clear that the low temperature intensity can be decomposed into the sum of three parts.

1) $I(\omega)^2 \left| A\hat{e}_i(2\omega) \chi^{EQ}_{ijkl}(\varphi) \hat{e}_j(\omega) \partial_k \hat{e}_l(\omega) \right|^2$

2) $I(\omega)^2 \left| A\hat{e}_i(2\omega) \chi^{EQ}_{ijkl}(\varphi) \hat{e}_j(\omega) \partial_k \hat{e}_l(\omega) \right| \left| A\hat{e}_i(2\omega) \chi^{ED}_{ijk}(\varphi, T) \hat{e}_j(\omega) \hat{e}_k(\omega) \right| \times 2 \cos \eta$

3) $I(\omega)^2 \left| A\hat{e}_i(2\omega) \chi^{ED}_{ijk}(\varphi, T) \hat{e}_j(\omega) \hat{e}_k(\omega) \right|^2$

This means that our measurement of the overall SHG intensity is sensitive to a sum of terms that scale differently with $\chi^{ED}_{ijk}(\varphi, T)$. Moreover we cannot directly determine the optical phase angle $\eta$ between the EQ and ED contributions to SHG in our experiments. Therefore an accurate measure of the temperature dependence of $\chi^{ED}_{ijk}(\varphi, T)$ cannot be made.

## S8. Landau free energy expansion for Néel and hidden order in $Sr_2IrO_4$

We perform a Landau free energy expansion in the Néel and hidden order parameters labeled by **S** and **Ω** respectively. The former represents the Néel vector of the dipolar antiferromagnetic ordered phase and the latter can represent either the in-plane toroidal moment or in-plane magnetic quadrupole moments[21] of the inversion symmetry broken hidden ordered phase of $Sr_2IrO_4$. Assuming that **S** and **Ω** break independent symmetries, the free energy F expanded out to quartic order in **S** and **Ω** is given by:

$$F = \alpha_N (T - T_N) (\mathbf{S} \cdot \mathbf{S}) + \beta_N (\mathbf{S} \cdot \mathbf{S})^2 + \alpha_\Omega (T - T_\Omega) (\mathbf{\Omega} \cdot \mathbf{\Omega}) + \beta_\Omega (\mathbf{\Omega} \cdot \mathbf{\Omega})^2 - \gamma (\mathbf{S} \cdot \mathbf{S})(\mathbf{\Omega} \cdot \mathbf{\Omega})$$

The free energy expansion is comprised simply of the free energy terms for each order parameter separately together with a fourth order coupling of the order parameters squared. There is no symmetry that enforces the critical temperatures $T_N$ and $T_\Omega$ to coincide. The fact that they are experimentally observed to be very close in $Sr_2IrO_4$ suggests a microscopic mechanism by which one can induce the other. One possibility is that loop-current order enhances the exchange coupling between dipolar moments on neighboring sites, which would mean that γ is positive. We note that since $T_N$ is driven by the development of *c*-axis coherence in $Sr_2IrO_4$[22], then if such a mechanism is at play it likely involves the hidden order enhancing interlayer super-exchange.

We note that generically two other terms can be constructed to quadratic order in **S** and **Ω** in the Landau free energy expansion. 1) $(\nabla \times \mathbf{S}) \cdot \mathbf{\Omega}$ : This term is what gives rise to toroidal order in spiral magnets like $LiCoPO_4$[23,24] where toroidal order is parasitic to the magnetic order. However we rule out this term because no spiral antiferromagnetic order ($\nabla \times \mathbf{S} = 0$) is observed in $Sr_2IrO_4$ from neutron and x-ray magnetic structure refinements. 2) $(\mathbf{S} \times \mathbf{P}) \cdot \mathbf{\Omega}$ : In principle an antiferroelectric order could combine with the antiferromagnetic order to yield a uniform toroidal order, where the antiferroelectric order consists of a staggered electric polarization **P** along the *c*-axis. However we rule out this term because no evidence of antiferroelectric ordering across $T_\Omega$ has been observed in dielectric measurements[25].

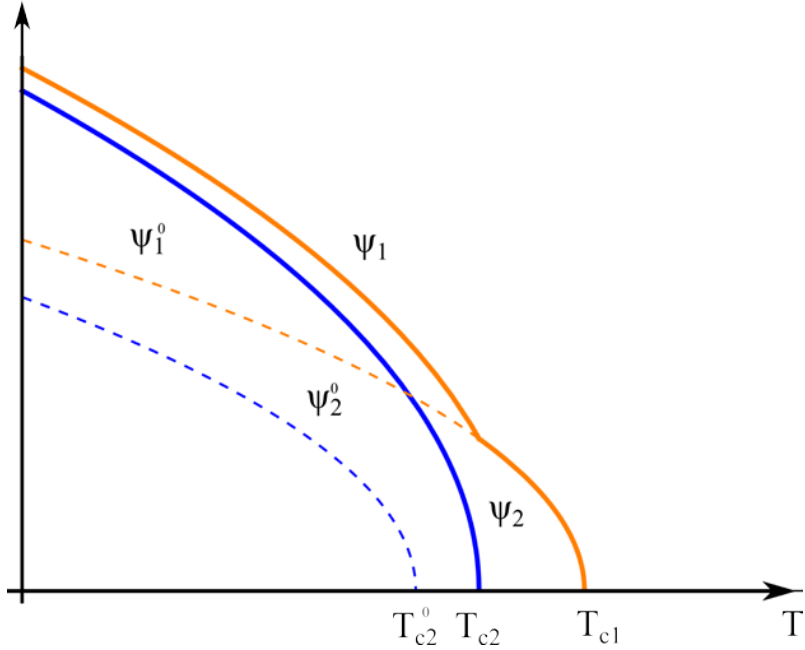

Fig. S7. The behavior of any two independent order parameters $\psi_1$ and $\psi_2$ (solid lines) with an attractive biquadratic interaction $\gamma > 0$. The non-interacting order parameter values $\psi_1^0$ and $\psi_2^0$ are given by the dashed lines. The attractive interaction can raise the lower transition temperature significantly ($T_{c2}^0 \to T_{c2}$), but will not generically lead to simultaneous transitions.

Our free energy expansion assumes that **S** and **Ω** break independent symmetries, which is the case for the Néel symmetry group 2/m1' and either of the two hidden order symmetry groups 2'/m and m1', all of which are irreducible representations of the high temperature crystallographic symmetry group 4/m1' (here, we have explicitly included the time-reversal symmetry, 1' in the group name). In such a case, the temperature dependence of the two order parameters **S** and **Ω** will have the generic form shown in Fig. S7, assuming a small attractive ($\gamma > 0$) biquadratic coupling between them.

# S9. Survey of $T_\Omega$ in $Sr_2Ir_{1-x}Rh_xO_4$ across spatial regions and across samples

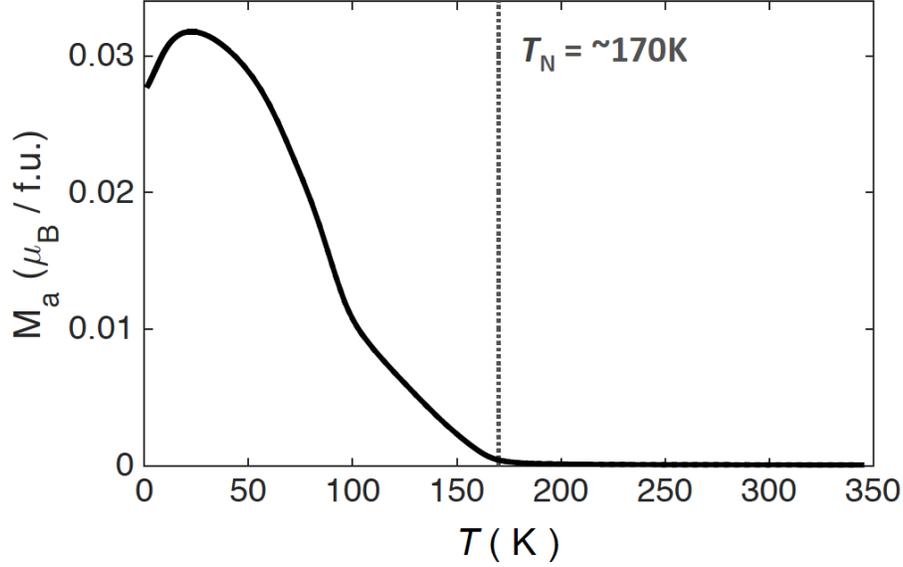

Fig. S8. The temperature dependence of the magnetization of $Sr_2Ir_{1-x}Rh_xO_4$ (x ~ 0.04) along the *a*-axis. The data were collected under an external magnetic field of $\mu_0 H = 0.1$ T. The onset of Néel order occurs at $T_N$ ~ 170 K, which is consistent with resonant x-ray diffraction data[24].

To confirm that the SHG curves measured from $Sr_2Ir_{1-x}Rh_xO_4$ shown in Fig. 4a-c do not arise from isolated regions of the sample with imperfect crystallographic or magnetic structure, we verified that our measurements are highly reproducible across different crystals, different locations within the crystals and even under different polarization geometries. We describe in detail a typical set of such checks on $Sr_2Ir_{0.96}Rh_{0.04}O_4$ in this section.

A typical temperature dependent magnetic susceptibility curve of $Sr_2Ir_{0.96}Rh_{0.04}O_4$ is shown in Fig. S8. There is a clear onset of Néel order at $T_N$ ~ 170 K, which is consistent with resonant x-ray diffraction measurements[20], and no features of any phase transition at higher temperatures. Figure S9 shows temperature dependent SHG intensity curves measured from two different pieces of $Sr_2Ir_{0.96}Rh_{0.04}O_4$ crystals on select locations under both $P_{in}$-$S_{out}$ and $S_{in}$-$S_{out}$ polarization geometries. The data consistently show that $T_\Omega$ ~ 220 K, which is 50 K higher than the onset of Néel order. The fact that below $T_\Omega$ the SHG intensity at $\varphi = 78°$ increases under $P_{in}$-$S_{out}$ and decreases under $S_{in}$-$S_{out}$ geometries is fully consistent with the RA-SHG data [see Fig. 1 and Fig. S1]. We also point out the absence of any observable kink feature in the temperature dependence curves at $T_N$, which indicates that the hidden order is not strongly coupled to the Néel order as predicted in Section S8. The absence of any observable kink at $T_N$ also re-affirms the expectation that our technique is not sensitive to any antiferromagnetic ordering in $Sr_2Ir_{1-x}Rh_xO_4$ owing to their centrosymmetric structure.

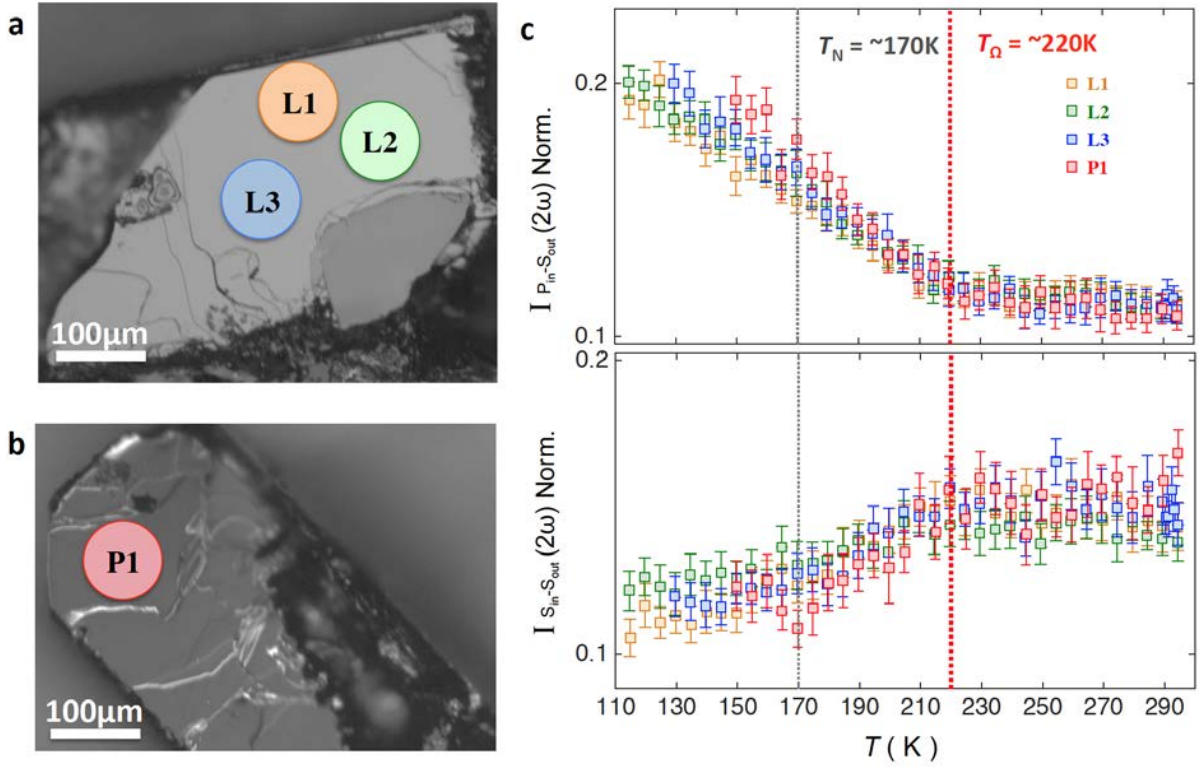

Fig. S9. Conventional optical microscope images of the cleaved surfaces of lightly hole-doped $Sr_2Ir_{1-x}Rh_xO_4$ (x ~0.04) crystals are shown in panels (a) and (b). RA-SHG patterns collected from locations L1 through L3 and P1 are displayed in this figure. (c) Temperature dependence of SHG intensity collected under $P_{in}$-$S_{out}$ and $S_{in}$-$S_{out}$ polarization geometries across different locations and crystals at $\varphi = 78°$. Data show a consistent value of $T_\Omega \sim$ 220 K, which is approximately 50 K higher than the onset of Néel order determined via magnetization measurements.

**Supplementary References**